\documentclass[runningheads]{svmult}

\usepackage{makeidx}   % allows index generation
\usepackage{graphicx}  % LaTeX graphics tool for including eps-figure files
\usepackage{subeqnar}  % subnumbers individual equations within an array
\usepackage{multicol}  % used for the two-column index
\usepackage{physprbb}  % modified textarea for proceedings,lecture notes, etc.
\makeindex             % used for the subject index
\usepackage{axodraw}   % Feynman graphs

\newcommand{\greeksym}[1]{{\usefont{U}{psy}{m}{n}#1}}

\newcommand{\uPi}{\mbox{\greeksym{P}}}
\newcommand{\be}{\begin{equation}}
\newcommand{\ee}{\end{equation}}
\newcommand{\bea}{\begin{eqnarray}}
\newcommand{\eea}{\end{eqnarray}}
\newcommand{\bsa}{\begin{subeqnarray}}
\newcommand{\esa}{\end{subeqnarray}}
\newcommand{\eps}{\varepsilon}
\newcommand{\Veff}%
{{\cal V}^{\mbox{\scriptsize p}}_{\mbox{\scriptsize eff}}}

\def\ra{\rightarrow}
\def\de{\Delta}
\def\al{\alpha}
\def\sig{\Sigma}
\def\gam{\Gamma}
\def\om{\omega}
\def\fdag{F^\dagger}
\def\ann{a_{nn}}
\def\kv{\vec{k}}
\def\epsk{\epsilon(\kv)}
\def\ll{\big\langle}
\def\rr{\big\rangle}
\def\sd{^3\!S\!D_1}
\def\pf{^3\!P\!F_2}
\def\ss{^1\!S_0}
\begin{document}

\title*{Superfluidity in Neutron Star Matter}

%\toctitle{Focusing of a Parallel Beam to Form a Point
%\protect\newline in the Particle Deflection Plane}
% allows explicit linebreak for the table of content
%\titlerunning{Focusing of a Parallel Beam}
% allows abbreviation of title, if the full title is too long
% to fit in the running head

\author{U. Lombardo\inst{1} \and H.-J. Schulze\inst{2}}

\institute{ 
 Dipartimento di Fisica, Universit\`a di Catania,\\
 Corso Italia 57, I-95129 Catania, Italy
\and
 Departament d'Estructura i Constituents de la Mat\`eria,
 Universitat de Barcelona,\\ Av. Diagonal 647, E-08028 Barcelona, Spain} 

\maketitle  

%\begin{abstract}
% The abstract\index{abstract} is optional. If present it should summarize
% the contents of the paper in at least 70 and at most 150 words; neither
% too long nor too short but to the point!
%\end{abstract}

%==============================================================================
\section{Introduction, General Formalism}

The research on the superfluidity of neutron matter can be traced back to 
Migdal's observation that neutron stars are good candidates for being 
macroscopic superfluid systems \cite{MIGD}. 
And, in fact, during more than two decades of 
neutron-star physics the presence of neutron and proton superfluid phases has 
been invoked to explain the dynamical and thermal evolution of a neutron star. 
The most striking evidence is given by post-glitch timing observations 
\cite{SAULS,PINE},
but also the cooling history is strongly influenced by the possible presence
of superfluid phases \cite{TSU,HEI}.
On the theoretical side, the onset of superfluidity in 
neutron matter or in the more general context of nuclear matter was 
investigated soon after the formulation of the 
Bardeen, Cooper, and Schrieffer (BCS) theory of superconductivity \cite{BCS} 
and the pairing theory in atomic nuclei \cite{BMP,BOHR}. 

The peculiar feature of a nucleon system is that it is a strongly interacting
Fermi system with a force which has a short-range repulsive component and a 
long-range attractive one.  
The first question raised by scientists was 
whether or not the strongly repulsive core might prevent the formation of a 
superfluid state. 
But it was indeed shown \cite{CMS} that the BCS approach, based on the 
mechanism of Cooper pairs, can be successfully extended to nuclear matter 
and that superfluid states could in fact exist for a wide class of 
nucleon--nucleon potentials \cite{ES}. 
The second question is related to the fact that the superfluid state of 
nuclear matter is a selfsustaining state in the sense that nucleons 
participating to the pairing coupling also screen the pairing itself.
From this point of view one expects the strong correlations to play an
important role in delimiting the magnitude of the pairing gap.    
Therefore it appears necessary to go beyond the pure BCS approach and 
properly add the effects of the medium polarization as self-energy and 
vertex corrections.

Since the neutron star ``laboratory'' can only provide an indirect evidence of 
the nucleon pairing in infinite matter and its relation with the pairing in 
nuclei is still too much model dependent, we need to rely on very accurate 
quantitative theoretical predictions of its properties such as energy gap, 
superfluidity density domain, critical temperature, and other physical 
quantities associated with the various superfluid states. 
These quantities may only be obtained from {\em ab initio}
calculations, i.e., microscopic approaches using as input the bare 
nucleon-nucleon interaction, because we are exploring a density domain much
wider than the saturation region where phenomenological interactions such as 
Skyrme forces are well suited. 
There is a more basic reason to refrain from
using effective interactions, that is a double counting of the
particle-particle (p-p) correlations incorporated in the effective
interaction, but also in the gap equation.

Fortunately, for more than two decades {\em realistic} potentials, 
based on field-theoretical approaches, 
have been supplied to describe the bare nucleon-nucleon interaction. 
The term `realistic' means that 
the parameters contained in such potentials are adjusted to simultaneously 
reproduce the experimental phase shifts of nucleon-nucleon scattering and 
the binding energies of the lightest nuclei. 

In this chapter the problem of superfluidity in neutron matter is surveyed 
and special emphasis is devoted to new theoretical developments and 
calculations. 
In the following section the general formalism of pairing in 
a strongly interacting Fermi system is presented. 
In Sect.~\ref{s:bcs} the possible superfluid states of nucleon matter are 
described within the BCS theory extended to non-zero angular momentum.  
In the last section some aspects of the generalized gap equation will
be discussed, including the medium polarization effects at very low density
(Sect.~\ref{s:ldl}), the induced interaction approach (Sect.~\ref{s:pol}),
and the role of self-energy corrections (Sect.~\ref{s:self}).

We mention three previous works for a comprehensive study of superfluidity in
nuclear matter: 
the early papers based on the generalized BCS-Bogolyubov
theory \cite{TAM,TAKA93}, which already give a systematic survey of most
superfluid states of nuclear matter; 
the second one \cite{CHEN93}, based on the method of the correlated basis 
functions, mainly focussed on the $\ss$ pairing, but
containing a wide discussion of the important medium correlation effects;
and lastly, a recent more general overview of pairing in 
nuclear matter \cite{ULOM}.

%==============================================================================
\subsection{Green Function Formalism, Generalized Gap Equation}
\label{s:green}

In this section we briefly review the main points of the treatment 
of a superfluid Fermi system within the Green function formalism. 
A detailed account is given in various 
textbooks \cite{ABRI,NOZ63,NOZ66,SCHRIEF,MIG,RS}.

\begin{figure}[t] %------------------------------------------------------------
\includegraphics[height=.24\textheight,bb=10 600 10 830]{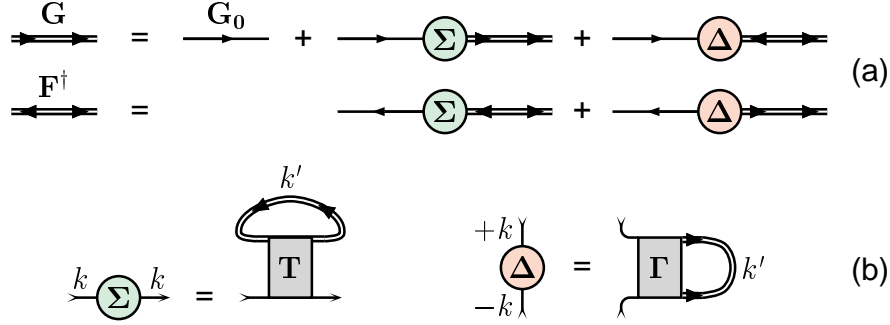} %-20
\caption[]
{({\bf a}) Diagrammatic representation of the Gorkov equations.
 ({\bf b}) Equations for the self-energy $\sig$ and gap function $\de$}
\label{f:gor}
\end{figure}

The principal equations describing a superfluid system are the Gorkov
equations, that can be considered a generalization of the Dyson equation 
for a normal Fermi system.
A diagrammatic representation of the Gorkov equations is shown in 
Fig.~\ref{f:gor}(a).
 %-----------------------------------------------------------------
They express the relation between normal and anomalous propagators 
$G$ and $F$, defined by
\bsa
 G(1,2) &=& {1\over\I} \ll T(\Psi_1 \Psi_2^\dagger) \rr  \:,
\\
 F(1,2) &=& {1\over\I} \ll T(\Psi_1 \Psi_2) \rr  \:,
\esa
and the self-energy $\sig$ and gap function $\de$.
In a homogeneous system, these four quantitites depend only on a 
four-vector $k=(k_0,\kv)$ and can be written as $2\times 2$ matrices in 
spin space, for example,
\be
  {\bf\de}(k) = \left( \begin{array}{cc} 
  \de_{\uparrow\uparrow} & \de_{\uparrow\downarrow} \\ 
   \de_{\downarrow\uparrow} & \de_{\downarrow\downarrow} 
                    \end{array} \right)(k) \:.
\ee
Using the free fermion propagator
\be
  G_0(k) = { e^{\I 0 k_0} \over k_0 - \kv^2\!/2m + \mu + \I 0 k_0}  \:,
\ee
and defining
\be
 \eps(k) = {\kv^2 \over 2m} + \Sigma(k_0,\kv) - \mu  \:,
\ee
one can write the system of equations explicitly as
\be
  \left( \begin{array}{cc}
  [k_0 - \eps(+k)]{\bf 1}  & {{\bf\de}(k)}            \\ 
  {{\bf\de\!^\dagger}(k)}    & [k_0 + \eps(-k)]{\bf 1}
         \end{array} \right)
  \left( \begin{array}{c} {{\bf G}(k)} \\ {{\bf\fdag}(k)} \end{array} \right) =
  \left( \begin{array}{c} {\bf 1} \\ {\bf 0} \end{array} \right) \:, 
\label{e:fg}
\ee
where $\bf 1$ denotes the two-dimensional unit matrix.
In order to take into account at the same time pairing correlations 
$\de_S$ with spin $S=0$ and $S=1$, one can make the ansatz
\be
  {\bf\de} = \left( \begin{array}{cc} 
  0 & +\de_0 + \I \de_1 \\ 
  -\de_0 + \I \de_1 & 0
                    \end{array} \right) \:,
\ee
%${{\bf\fdag}}$ has the same structure, 
and equivalently for $\bf\fdag$,
whereas the self-energy $\bf\sig$ and $\bf G$ are diagonal in the spin indices.
If the ground state is assumed to be time-reversal invariant, the gap
function has in general the structure of a 
unitary triplet state \cite{BALDO1,BALDO2},
i.e., it fulfills
\be
 {{\bf\de\!^\dagger}(k)} {\bf\de}(k) = \de(k)^2 \bf 1  \:,
\ee
where by $\Delta(k)^2$ we denote the determinant of ${\bf\de}$ in spin space. 

The system Eq.~(\ref{e:fg}) can then be inverted with the solution
\bsa
 G(k) &=& {k_0 + \eps(-k) \over D(k)} \:, 
\label{e:gg}
\\
 \fdag_S(k) &=& {\Delta_S(k) \over D(k)} \:,
\esa
where
\be
 D(k) = \left[ k_0 - \eps(+k) \right]  \left[ k_0 + \eps(-k) \right] 
         - \Delta_0(k)^2 - \Delta_1(k)^2  \:,
\label{e:dd}
\ee
that expresses the propagators $G$ and $\fdag$ in terms of $\sig$ and $\de$, 
respectively.

In order to determine uniquely the four quantities
one needs two more equations, 
which relate $\sig$ and $\de$ to the interaction.
These equations are displayed in Fig.~\ref{f:gor}(b) and read explicitly
\bsa
  \sig_{\al\al}(k) &=& {1\over\I} \int\! {{\rm d}^4 k' \over (2\pi)^4}
  \sum_\beta \ll k\al,k'\beta |T| k\al,k'\beta \rr  \, G_{\beta\beta}(k') \:,
\label{e:sga}
\\
  \de_{\al\beta}(k) &=& \,{\I} \int\! {{\rm d}^4 k' \over (2\pi)^4}
  \sum_{\al',\beta'} \ll k\al,-k\beta |\gam| k'\al',-k'\beta' \rr  
  \,F_{\al'\beta'}(k') \:,
\label{e:sgb}
\esa
where greek letters denote spin indices and
$T$ and $\gam$ are the scattering matrix
and the irreducible interaction kernel, respectively.
Clearly these equations cannot be solved in full generality, 
but one has to recur to some approximation at this stage.
The simplest, very common, BCS approximation, is to replace $T$ and $\Gamma$
by the leading term, namely the bare interaction $V$.
In this case the interaction is energy independent and the 
$k_0$ integration in Eqs.~(\ref{e:sga},\ref{e:sgb}) can be carried out 
trivially, leading to
\bsa
 \Sigma(\kv) &=& \sum_{\kv'} { v_{\kv'}^2 \over 2 } 
 %\ll \kv,\kv' |V| \kv,\kv' \rr_a    
 \Big[ \ll\kv,\kv'|{V_0+3V_1}|\kv,\kv'\rr 
\nonumber\\[-4mm]&&\hskip14mm 
 - \ll\kv,\kv'|{3V_1-V_0}|\kv',\kv\rr \Big] \,\;   
 \ ,\quad 
 v^2 = {1\over2}\left( 1 - {\eps\over E} \right) \:,
\\
 \Delta_S(\kv) &=& \sum_{\kv'} (u_S v)_{\kv'} 
 \ll \!\!+\!\kv',-\kv' |V_S| \!+\!\kv,-\kv \rr_a 
 \ ,\quad
 u_S v = {-\de_S\over 2E} \:,
\esa
where
\be
 E^2 = \eps^2 + \de_0^2  + \de_1^2  
 \ , \quad
 \eps = {\kv^2\over2m} + \sig(\kv) - \mu  \:.
\ee
Together with an equation fixing the chemical potential $\mu$ 
for given density $\rho$,
\be
 \rho = 2\sum_{\kv} v_{\kv}^2 \:,
\ee
this is the coupled set of equations that needs to be solved in order
to find the Hartree-Fock self-energy $\sig_{\rm HF}(\kv)$ and the 
BCS gap function $\de_{\rm BCS}(\kv)$ in a superfluid system.
In contrast to a normal Fermi system, the smooth occupation 
numbers $v_{\kv}^2$ instead of the Fermi function $\theta(k_F-|\kv|)$ 
appear in the HF equation.

%==============================================================================
\section{BCS Approximation}
\label{s:bcs}

In this section we present the solutions of the BCS gap equation
\bsa
 \de_{TS}(\kv) &=& - \sum_{\kv'}  \ll \kv | V_{TS} | \kv' \rr
 {\de_{TS}(\kv')\over 2E(\kv') }  \:,
\label{e:gap1}
\\
 \rho = {k_F^3\over 3\pi^2} 
  &=& 2 \sum_{\kv} {1\over2} \left[ 1 - {\epsk \over E(\kv)} \right] \:,
\label{e:rho}
\esa
where 
\be
  E(\kv)^2 = \epsk^2 + \!\!\sum_{T,S=0,1}\!\!\!\de_{TS}(\kv)^2 
\ , \quad
 \epsk = e(\kv) - \mu
\label{e:ene}
\ee
with $\mu$ being the chemical potential 
and $e(\kv)$ the single-particle spectrum. 
Different realistic nucleon-nucleon potentials $V$ 
\cite{PARIS,V14,MACH89,NIJ,V18,BONN}
will be used as input.
The equations are valid for pure neutron matter ($T=1$) 
and also for {\em symmetric} nuclear matter ($T=0,1$),
having extended the derivation in the previous section by including the 
isospin quantum number $T$ in analogy to the spin $S$.

%==============================================================================
\subsection{Pairing in Different Partial Waves}

In order to reduce the three-dimensional integral equation (\ref{e:gap1}) 
to a set of one-dimensional ones,
it is advantageous to perform partial wave expansions of
the potential and the gap function. 
In this way one arrives at separate equations
in the different $(TSLL')$ channels of the interaction,
provided an angle-average approximation is made by replacing
$\de(\kv)^2 \ra \int\!{\D\vec{\hat{k}}/4\pi}\, \de(\kv)^2$
in Eq.~(\ref{e:ene}).
The following equations for the partial wave components of the gap function 
are then obtained \cite{TAKA93,AMU285,BCLL92,ELGA96,BEEHS98,KHODEL98}:
\be
  \Delta_{TSL}(k) = - {1\over\pi} \int_0^\infty \!\! \D k' k'^2 \sum_{L'}
  {V^{TS}_{LL'}(k,k') \over \sqrt{ \epsilon(k')^2 + \de(k')^2 }}
  \Delta_{TSL'}(k') \:,
\label{e:gap2}
\ee
where 
\be
 \de(k)^2 = \sum_{T,S,L} \de_{TSL}(k)^2  \:,
\ee
and with the matrix elements of the bare potential in momentum space
\be
  V^{TS}_{LL'}(k,k') =  
  \int_0^\infty \!\! \D r\, r^2\, j_{L'}(k'r)\, V^{TS}_{LL'}(r)\, j_L(kr) \:.
\label{e:v}
\ee
It should be noted that the different equations are still coupled due to the 
fact that the total gap appearing in the denominator 
on the r.h.s.~of Eq.~(\ref{e:gap2})
is the r.m.s. value of the gaps in the different partial waves.
The gap equation allows in principle the coexistence of pairing correlations 
with different quantum numbers ($TS$),
even though the different $(TS)$ channels are 
not mixed by the interaction.
In practice, however, so far no such mixed solutions
of the gap equation have been found: 
even if at a given density two or more uncoupled
solutions exist, the strong nonlinear character of the gap equation
prohibits a coupled solution.
(In finite nuclei such mixed solutions seem to exist under certain 
conditions \cite{AG}).
This means that in practice this $(TS)$-coupling can be neglected and that
at a given density the solution of the uncoupled gap equation with the 
largest gap is the energetically favoured one.

The only case when it is clearly necessary to keep the coupled equations
is the mixing of partial waves due to the tensor potential.
In this case the gap equation can be written in matrix form 
(for given $S=1,T,L$):
\be
 \left( \begin{array}{l} \Delta_L \\ \Delta_{L+2} \end{array} \right)\!(k) =
 - {1\over\pi} \int_0^\infty \!\!\D k' k'^2 {1\over E(k')}
 \left( \begin{array}{ll}
  V_{L,L} & V_{L,L+2} \\ V_{L+2,L} & V_{L+2,L+2}
 \end{array} \right)\!(k,k')
 \left(\begin{array}{l} \Delta_L \\ \Delta_{L+2} \end{array}\right)\!(k') 
\ee
with 
\be
 E(k)^2 = [e(k)-\mu]^2 + \Delta_L(k)^2 + \Delta_{L+2}(k)^2 \:.
\ee
This is relevant equation for the $\sd$ $(T=0)$ and $\pf$ $(T=1)$ 
channels, for example.

Let us finally mention that usually, apart from the $\sd$ channel,
the two equations (\ref{e:gap1}) and (\ref{e:rho}) can be decoupled 
by setting $\mu=e(k_F)$. 
The reason is the small value of the ratio $\de/\mu$,
so that a Fermi surface is still quite well defined.

\begin{figure}[b] %------------------------------------------------------------
\includegraphics[height=.41\textheight,bb=40 380 40 720]{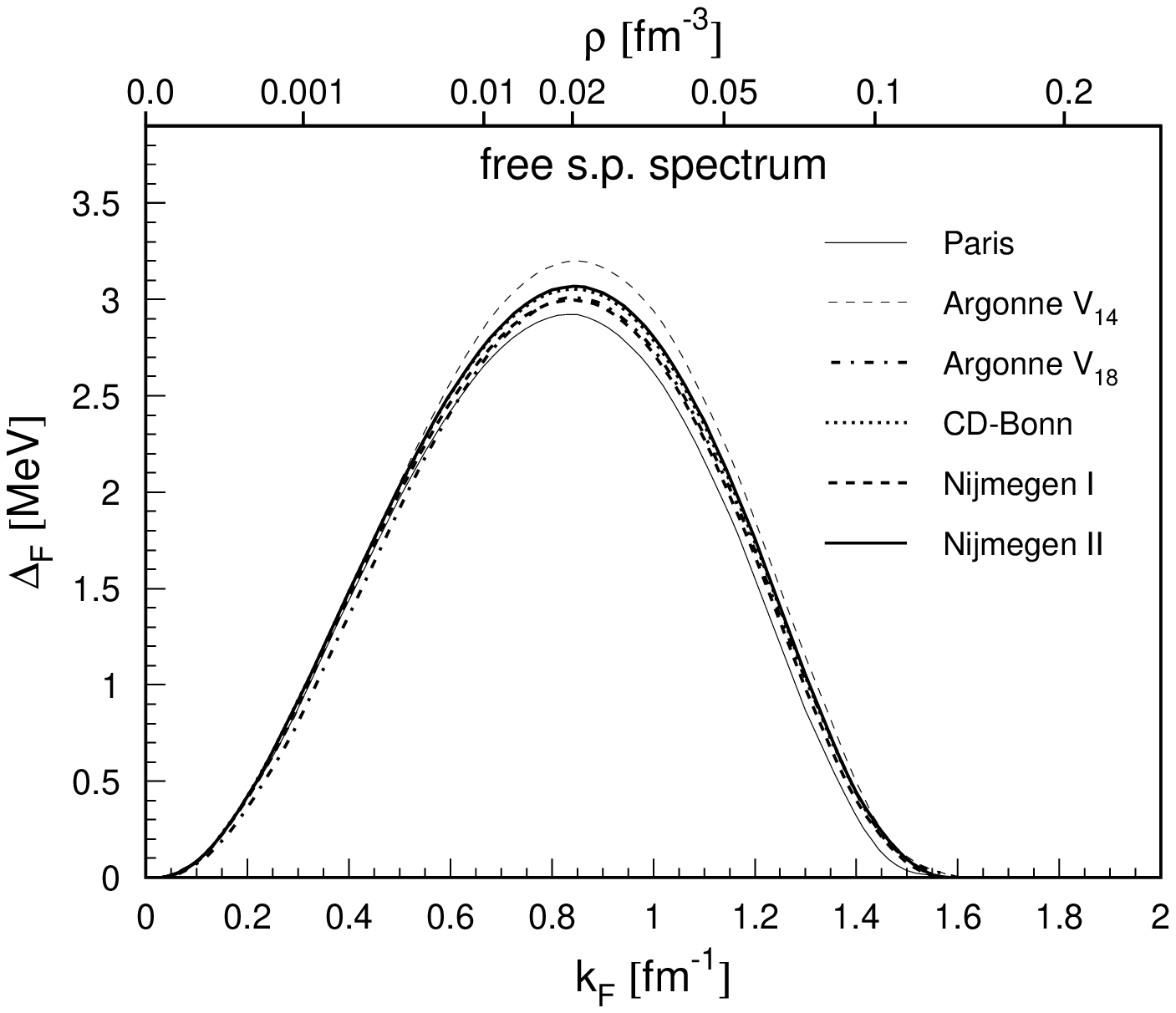}
\caption[]
{$\ss$ gap evaluated in BCS approximation with free single-particle spectrum 
 and different potentials}
\label{f:1s0}
\end{figure}

\begin{figure}[t] %------------------------------------------------------------
\includegraphics[height=.41\textheight,bb=40 380 40 720]{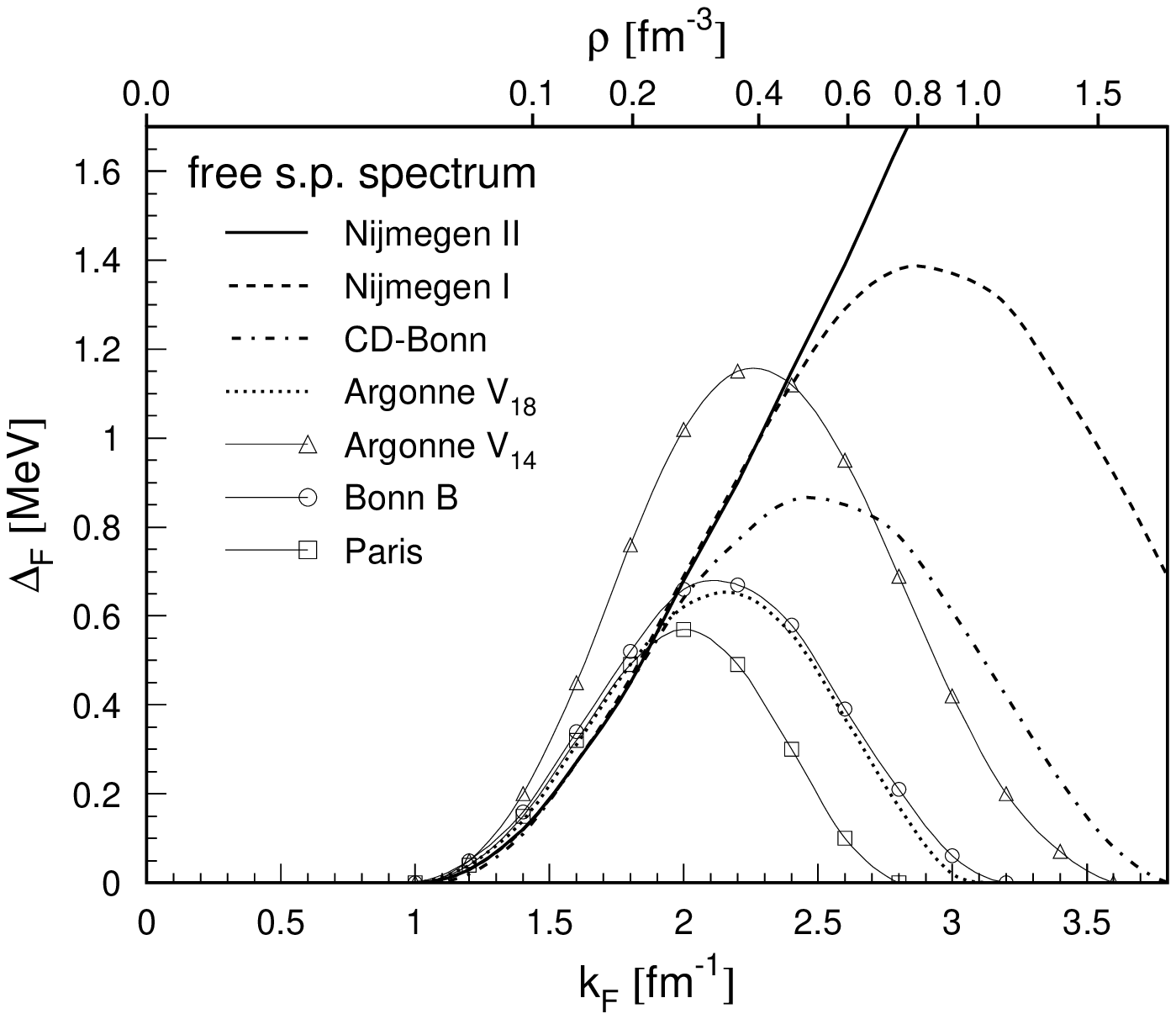}
\caption[]
{$\pf$ gap evaluated in BCS approximation with free single-particle spectrum 
and different potentials}
\label{f:3p2}
\end{figure}

We come now to the presentation of the results that are obtained by solving
the previous equations numerically,
using a kinetic energy spectrum $e(k)=k^2\!/2m$ for the moment.
In practice, one finds in pure neutron matter ($T=1$) gaps only in the
$\ss$ \cite{TAKA93,CHEN93,AMU85,BCLL90,KHODEL96,ELGA98} 
and $\pf$ \cite{TAKA93,AMU285,BCLL92,ELGA96,BEEHS98,KHODEL98} 
partial waves.
They are reported in Figs.~\ref{f:1s0} and \ref{f:3p2}, respectively.
 %-----------------------------------------------------------------
 %-----------------------------------------------------------------
It can be observed that the maximum pairing gap is about $3\,$MeV in the 
$\ss$ channel and of the order of $1\,$MeV in the $\pf$ wave.
It is remarkable that solutions obtained with different nucleon-nucleon 
potentials are nearly indistinguishable in the $\ss$ case, 
whereas in the $\pf$ wave such good agreement can only be observed up to
$k_F \approx 2\,\rm fm^{-1}$
(with the exception of the Argonne $V_{14}$ potential that is not very well 
fitted to the phase shifts),
from where on the predictions start to diverge from each other.
The reason \cite{BEEHS98,ELGA98} 
is the fact that the various potentials are constrained by the
phase shifts only up to scattering energies $E_{\rm lab}$ of about $350\,$MeV, 
which roughly corresponds to a Fermi momentum of 
$k_F\approx \sqrt{mE_{\rm lab}/2} \approx 2\,\rm fm^{-1}$.
Thus even on the BCS level the gap in the $\pf$ channel
at a neutron density higher than $\approx 0.3\,\rm fm^{-3}$ 
is at the moment not known.
Apart from that it is clear that the BCS approximation is not reliable
at the very large densities for which a gap is predicted in Fig.~\ref{f:3p2}.
However, nobody has so far attempted to include polarization effects in 
this channel.

Let us mention here for completeness that in symmetric nuclear matter
one finds very strong pairing of the order of $10\,$MeV in the 
$\sd (T=0)$ channel, reminiscent of the deuteron bound state
\cite{TAKA93,BALDO1,BALDO2,ALM1,VONDER,ALM2,SSALR95,ALM3,%
BALDO3,MORTEN,LOMBARDO},
and also a gap of the order of $1\,$MeV in the $^3\!D_2$ wave 
\cite{TAKA93,SED95}.
This, however, is probably not very relevant for neutron star physics, 
since a prerequisite for this $T=0$ neutron-proton pairing to take place
is the existence of (nearly) isospin symmetric nuclear matter,
the pairing correlations being rapidly destroyed by increasing 
asymmetry \cite{SAL97,ROEPKE00,SL00}. 
We will therefore not discuss this type of pairing further on. 
%This will be discussed in more detail in the following section.

%==============================================================================
\subsection{Pairing Gaps in Neutron Star Matter}
 
\begin{figure}[t] %-----------------------------------------------------------
\includegraphics[height=.4\textheight,bb=40 380 40 700]{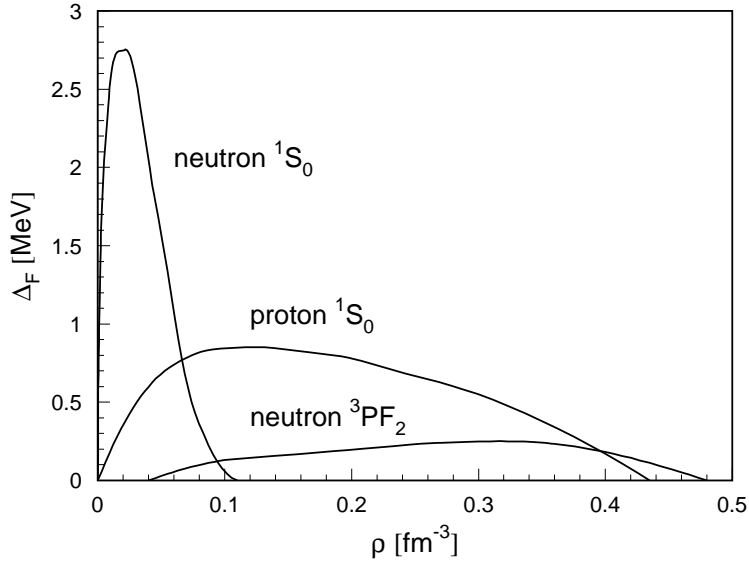}
\caption[]
{The different $T=1$ gaps in neutron star matter as a function of 
total nucleonic density}
\label{f:gapn}
\end{figure}

Let us now come to the $T=1$ gaps that can be expected in isospin asymmetric 
(beta-stable and charge neutral) neutron star matter.
On the BCS level, the only influence of isospin asymmetry on these gaps is via
the neutron single-particle energy $e(k)$ appearing in the gap equation.
Calulations have been performed using $e(k)$ determined in the 
BHF approximation extended to asymmetric nuclear matter 
\cite{BCLL92,ELGA96,CDL,BOMBA,BOMBA2,EEHE96,BBS,VPREH}, 
and typical results are shown in Fig.~\ref{f:gapn}.
 %-----------------------------------------------------------------
One observes that $\ss$ pairing can take place independently in the 
neutron and in the proton component of the matter.
If plotted as a function of total baryon density, the neutron pairing
occurs naturally at lower density and with a larger amplitude than
the proton pairing, because at the higher density the proton effective
mass is smaller and the pairing therefore more reduced.  
The same is true for the $\pf$ pairing in the neutron component, which is 
strongly reduced with respect to the calculation with a free spectrum 
shown in Fig.~\ref{f:3p2} above.
We stress that the results displayed in Fig.~\ref{f:gapn} can only be 
qualitative, because they clearly depend on the details 
(in particular the proton fraction)
of the equation of state that is used.
The results shown were obtained with a BHF EOS based on the Argonne $V_{14}$
potential and involving $n$, $p$, $e$, $\mu$ components \cite{BOMBA}.

%==============================================================================
\section{Beyond BCS}

In the previous section we have presented many results that were all obtained 
within the BCS approximation. 
%i.e., using only the first-order bare potential in the kernel of the 
%gap equation.
However, as has been explained in the introduction, 
%the BCS 
this
approximation amounts to a mean-field approach, equivalent to and consistent 
with the Hartree-Fock approximation in a normal Fermi system.
More precisely, the BCS approximation neglects completely 
any contribution beyond the bare potential to the interaction kernel $\gam$
%and the single-particle self-energy $\sig$
appearing in the general gap equation (\ref{e:sgb}).

Going consistently beyond the BCS approximation is however a very difficult
task and has been only partially achieved so far.
We will review in the following sections some aspects of these extensions.
We begin with a discussion of the situation at extremely low density, where
certain analytical results are known.
Following that, the general framework at more relevant densities will be 
set up, and we will briefly present the results that have been obtained 
so far by various authors.
Finally, in the last section, we will focus on a certain part of the 
problem that has recently been tackled, namely the treatment of  
the energy dependence of the self-energy that appears in the gap equation.

%==============================================================================
\subsection{Low Density}
\label{s:ldl}

In order to derive an exact analytical result for the pairing gap 
including polarization effects
that is valid at very low density
(more precisely, for $k_F<\!\!<1/|a|$, where $a$ is the relevant 
scattering length),
we begin again with the BCS gap equation,
\be
  \de_{k} = - \sum_{k'} V_{kk'} {1\over 2E_{k'}} \de_{k'} 
  \ ,\quad
  E_k = \sqrt{ (e_k-e_F)^2 + \de_k^2 }  \:.
\ee
It is then useful \cite{BCLL90,EEHE96,ANDER} to introduce a modified 
interaction $T$ that is given by the solution of the integral equation
\be
 T_{kk'} = V_{kk'}  - \sum_{k''}  T_{kk''} F_{k''} V_{k''k'} \:,
\ee
where $F_k$ is for the moment an arbitrary function.
(We have used the symbol $T$, although in general this quantity is
not to be identified with the scattering matrix).
Making use of this equation, the gap equation is transformed into
\be
  \de_{k} = - \sum_{k'} T_{kk'} 
  \left( {1\over 2E_{k'}} - F_{k'} \right) \de_{k'} \:.
\ee
Of particular interest is now the choice 
$F_k={\rm sgn}(k-k_F)/2E_k$, which leads to the set of equations
\bea
  \de_{k} &=& - 2 \sum_{k'<k_F} T_{kk'} {1\over 2E_{k'}} \de_{k'} \:,
\label{e:gapc}
\\
  T_{kk'} &=& V_{kk'}  
  - \sum_{k''} T_{kk''} {{\rm sgn}(k''-k_F)\over 2E_{k''}} V_{k''k'} \:.
\eea
Therefore, in the limit $\Delta/e_F\ra0$ that is approached with vanishing
density, $T$ does become identical to the free scattering matrix,
because $E_k \ra |e_k-e_F|$ in this situation.
At the same time, in the gap equation (\ref{e:gapc}) the interaction $T$ is 
now cut off at $k'=k_F$ 
(at the cost of introducing a factor 2), 
so that with vanishing density it is ultimately sufficient to use the 
low-energy result \cite{FW} for the $T$-matrix:
\be
  T_{kk'} \ra T_{00} = {4\pi \ann \over m} \:,   
\label{e:tlow}
\ee
where $\ann = -18.8{\,\rm fm}$ is the neutron-neutron scattering length.
This yields finally the gap equation
\be
  1 = -{4 k_F \ann \over \pi} 
  \int_0^1\! \D x\, {x^2 \over \sqrt{(1-x^2)^2 + (\de/e_F)^2} } \:,
\ee
which in the limit $\de/e_F \ra 0$ is solved by 
\cite{KHODEL96,GMB,BERTSCH,PET}
\begin{eqnarray}
 \Delta(k_F)
 \ {\buildrel k_F\ra 0  \over \longrightarrow}\ 
 \Delta_0(k_F) = 
 {8 \over \E^2} {k_F^2\over 2m} \exp\!\left[{\pi\over 2k_F \ann}\right] \:.
\label{e:appbcs}
\end{eqnarray}
This is the universal asymptotic result 
for the BCS pairing gap in a low-density Fermi system with negative 
scattering length.
Unfortunately its validity is limited to the region
$k_F<\!\!<1/|\ann| \approx 0.05\,\rm fm^{-1}$,
far below the densities of interest for neutron star physics
or even pairing in finite nuclei.

\begin{figure}[b] %------------------------------------------------------------
\includegraphics[height=0.14\textheight,bb=20 630 20 750]{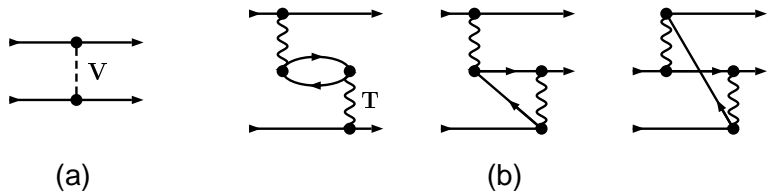}
\caption[]
{({\bf a}) Bare potential and 
({\bf b}) first-order (direct and exchange) polarization diagrams 
contributing to the interaction kernel in the low-density limit}
\label{f:pollow}
\end{figure}

Going now beyond the BCS approximation, in the low-density limit one 
should take into account the corrections to the interaction kernel
that are of leading order in density. 
Diagramatically these are the polarization diagrams of first
order (i.e., comprising one polarization ``bubble'') that are
displayed in Fig.~\ref{f:pollow}.
The interaction appearing in these diagrams is in the present case
the free scattering matrix $T$. 
 %-----------------------------------------------------------------
It can be shown \cite{GMB} that in the low-density limit it is again 
sufficient to 
neglect the momentum dependence of the $T$-matrix, as in Eq.~(\ref{e:tlow}).
One obtains then for the lowest-order polarization interaction
\be
 W_{\ss}(k,k') = -{ T_{00}^2 \over 8\pi} {1\over 2 k k'}
 \int_{|k-k'|}^{k+k'}\! \D q\, q\, \uPi(q) \:,
\ee
where 
\bea
  %4 \sum_h {f_h \fb_{h+q}\over e_h - e_{h+q} } = 
  \uPi(q) &=& 
  -{m k_F \over \pi^2}
  \left[{1\over2} + {1-x^2\over 4x}\ln\left|{1+x\over 1-x}\right| \right] 
  \ ,\quad 
  x = {q\over 2k_F} 
\label{e:pi}
\eea
%and where $N=m k_F/\pi^2$ is the density of states at the Fermi surface 
%for neutron matter.
is the static Lindhard function \cite{FW}.
Thus the lowest-order polarization modifies the 
BCS interaction kernel Eq.~(\ref{e:v})
by a (repulsive) term proportional to $k_F$, 
whereas any other polarization diagram 
contributes only in higher order of $k_F$.

We can use this result and insert it into the previously obtained approximation
in the BCS case:
\begin{eqnarray}
 \Delta(k_F)
 \ {\buildrel k_F\ra 0  \over \longrightarrow}\ 
 {8 \over \E^2} {k_F^2\over 2m} 
 \exp\!\left[ {\pi/2  \over \kappa + c\kappa^2 }\right]
 \ , \quad \kappa = k_F\ann  \:,
\label{e:}
\end{eqnarray}
where 
\begin{eqnarray}
 c &=& -{2\pi\over m k_F} 
 \int_0^{2k_F} \!{\D q\, q\over 2k_F^2} \,\uPi(q) 
 %\uPi(k_F,k_F)
 = {2\over3\pi} \left( 1+2\ln 2 \right) \approx 0.506
\end{eqnarray}
accounts for the polarization effects to first order.
Expanding now the argument of the exponential up to second order in $\kappa$,
one obtains for the ratio relative to the BCS value, Eq.~(\ref{e:appbcs}),
\begin{eqnarray}
 { \Delta(k_F) \over \Delta_0(k_F) }
 &=&
 \exp\!\left[ -{\pi\over2}c 
     \left[ 1 - c\kappa + {\cal O}(\kappa^2) \right]
     \right] 
\\
 &\approx&
 \left[ {1 \over (4\E)^{1/3}} \right]^{(1-c\kappa)}
\\
 &{\buildrel k_F\ra 0  \over \longrightarrow}& 
 {1 \over (4\E)^{1/3}} \:.
\label{e:apppol}
\end{eqnarray}
Let us stress that the above ``derivation'' of this result can only be 
considered heuristic. A rigorous proof was given originally 
in Ref.~\cite{GMB}.

Therefore, one arrives at the striking conclusion that in the low-density 
limit the polarization corrections suppress the BCS gap by a 
factor $(4\E)^{-1/3}\approx$ 0.45,
independent of the strength of the interaction $\ann$.
This is quite surprising, since the polarization interaction clearly vanishes 
with vanishing density; its effect on the pairing gap, however, does not.
The reason is the nonanalytical dependence of the gap on the interaction
strength, as expressed by Eq.~(\ref{e:appbcs}).

All this means that the BCS approximation cannot even be trusted at 
very low density, and that one can in general expect quite strong 
modifications due to polarization effects at higher density as well.

This low-density behaviour of neutron pairing could be met in the study 
of exotic nuclei with a long density tail or in nuclei embedded in
a neutron matter environment, as occurring in the neutron star crust.

%==============================================================================
\subsection{General Polarization Effects}
\label{s:pol}

To go beyond the simple low-density approximations derived before
requires considerable effort and has in fact so far not been accomplished in 
a satisfactory manner, so that ultimate results cannot be presented here.
The reason is that many effects that could be neglected in the low-density 
limit become important now, and that on the other hand the pairing gap
is extremely sensitive to even slight changes of the interaction kernel.

\begin{figure}[h] %------------------------------------------------------------
\includegraphics[height=0.45\textheight,bb=-0 410 -0 820]{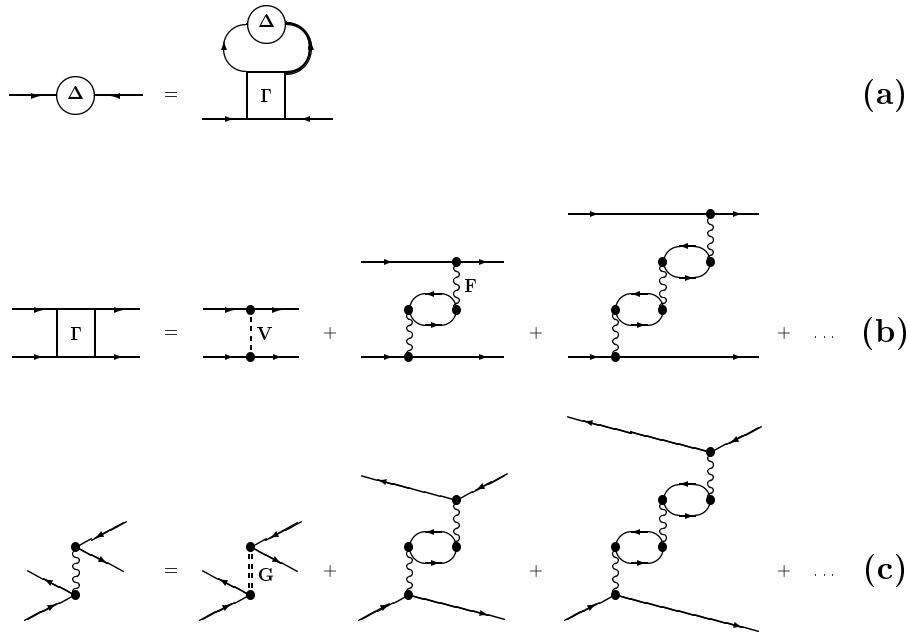}
\caption[]{
Determination of the interaction kernel $\Gamma$ in the gap equation ({\bf a}):
Polarization diagrams appear in the particle-particle channel ({\bf b}) 
as well as in the particle-hole channel ({\bf c}).
The leading diagrams in these channels   
are the bare potential $V$ (dashed line) and the $G$-matrix (double-dashed
line), respectively}
\label{f:pol}
\end{figure}

First, 
outside the low-density region $k_F<\!\!<1/|\ann|$,
the interaction kernel has to be extended beyond the 
lowest-order polarization diagrams. 
This means summing up polarization diagrams of all orders in the 
particle-particle channel, but also including them in the particle-hole
channel, replacing the $T$-matrix by the general particle-hole
interaction $F$.
In this way a self-consistent scheme is established that is depicted 
in Fig.~\ref{f:pol}.
 %-----------------------------------------------------------------
It requires as input the Brueckner $G$-matrix and yields ideally the
interactions in the particle-particle as well as particle-hole channel,
$\Gamma$ and $F$, respectively.
It is clear that in practice an exact solution is impossible, but that usually
strong approximations have to be performed that cast a doubt on the 
reliability of the results that are obtained.
We will later discuss this point in some more detail.

Second, related to the previous item, the energy dependence 
of the full gap equation [see Eq.~(\ref{e:sgb})] needs to be 
taken into account.
This is obvious, since the interaction kernel becomes now a complex,
energy-dependent quantity.
To our knowledge, this problem has so far not been studied in detail 
in the literature. 
It is therefore not known how far the gap could be changed by this
more elaborate treatment of the equations.

Third, and in connection with the two previous points,
the choice of a particular interaction kernel 
requires also the choice of a compatible self-energy appearing in the 
gap equation.
This will be explained in more detail in the following section, as it has
recently been addressed in the literature.

\begin{figure}[b] %------------------------------------------------------------
\includegraphics[height=.4\textheight,bb=40 380 40 690]{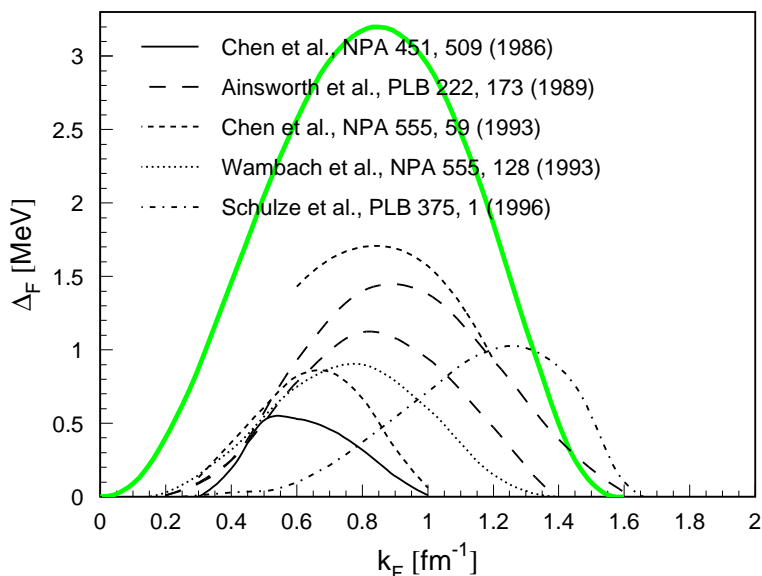}
\caption[]
{The $\ss$ gap in pure neutron matter predicted in several publications
taking account of polarization effects.
The curve in the background shows the BCS result}
\label{f:polgap}
\end{figure}

It should be clear by now that the influence of medium effects on pairing
constitutes an extremely difficult problem.
Consequently the results that can be found in the literature 
\cite{CHEN93,CLARK1,CHEN86,AINS89,WAMBACH,SCHU}
addressing this task in certain approximations agree only on the fact that
generally a strong reduction with respect to the BCS gap is obtained.
A collection of these results is displayed in Fig.~\ref{f:polgap}. 
It can be seen that the precise amount and density dependence of the 
suppression vary substantially between the different approaches 
and must be considered unknown for the time being.
 %-----------------------------------------------------------------

The most advanced description of medium polarization effects is based on the
Babu-Brown induced interaction model 
\cite{BABU,SJO73,BACK73,JACK82,DICK83,BACK85}. 
The microscopic derivation 
of the effective interaction starts from the following physical idea: 
The particle-hole (p-h) interaction can be considered as made of a 
{\em direct} component containing the short-range correlations and an 
{\em induced} component due to the exchange of the collective excitations 
of the medium.

Let us consider a homogeneous system of fermions interacting via an 
instantaneous potential $V$, which is also translationally and rotationally
invariant. 
Collective excitations are described by the ring series, which 
can easily be summed up \cite{FW}, and the p-h interaction can be written as
(disregarding for the moment spin degrees of freedom)
\be
 V_{\rm ph}(q) = V(q) + \frac{\uPi(q) V(q)^2}{1 - \uPi(q) V(q)}
\label{e:vind}  \:.
\ee
In this simple case the interaction itself plays the role of the direct term 
and the sum of the ring series that of the induced term.

In the nuclear case the presence of the hard core imposes the bare
interaction $V$ to be renormalized in order to incorporate the 
short-range correlations. 
This goal is reached by introducing the $G$-matrix,
which sums particle-particle (p-p) ladder diagrams to all orders, and the 
diagrammatic expansion can be recast just replacing $V$ by $G$. 
But now the new ring series cannot be summed up any longer, mainly
since the $G$-matrix is nonlocal. 
An averaging procedure 
%on the Fermi sphere 
has been devised to bring 
the $G$-matrix into a local form $G(q)$ \cite{SCHU}. 
Then, in analogy to Eq.~(\ref{e:vind}), the p-h interaction $F$ 
is given by
\be
 F(q) = G(q) + \frac{\uPi(q) G(q)^2}{1 - \uPi(q) G(q)}
\label{e:gind}  \:.
\ee   
In a simplified version of the theory the direct term now coincides with 
the $G$-matrix and the induced term is the approximate sum of the 
renormalized ring diagrams. 
In Fig.~\ref{f:pol} the direct term is represented by the first diagram 
on the r.h.s.~of the series (c). 
The next diagrams form the ring (or bubble) series. 

But, since the RPA series  
with the $G$-matrix produces a too strong polarizability of nuclear matter, 
in the Babu-Brown approach it has been proposed to include in the RPA series 
the full p-h interaction itself, since the particle (hole) coupling 
vertex with the p-h bubble can indeed be identified with the 
irreducible p-h interaction. 
The series (c) of Fig.~\ref{f:pol} is, in fact, the final result of such 
a procedure, where the wiggles represent the effective p-h interaction 
$F$ on either side.
If we denote by $F_d$ the direct interaction ($G$-matrix in our case) 
and by $F_i$ the induced interaction, the effective p-h interaction can be
written in the form (at the Landau limit, in which $\vec p_1$ and $\vec p_2$ 
are restricted to the Fermi surface) 
\be
 F(\vec p_1,\vec p_2) = 
 F_d(\vec p_1,\vec p_2) +  F_i(\vec p_1,\vec p_2;F)  \:,  
\label{e:find}
\ee
where $F_i$ is, as said before, the RPA series of Fig.~\ref{f:pol}(c) 
with the $G$-matrix replaced by $F$ itself. 
The previous equation clearly entails a self-consistent 
procedure to determine the interaction $F(\vec p_1,\vec p_2)$.  
Referring to \cite{SCHU} for details, one may reduce Eq.~(\ref{e:find}) to
a numerically tractable form 
\be
 F(q) = G(q) + \frac{\uPi(q) F(q)^2}{1 - \uPi(q) F(q)}  \:,
\label{e:wind}
\ee
which corresponds to replacing the $G$-matrix by $F$ in the induced term.   

Once the irreducible p-h interaction $F$ has been determined, 
one can construct the irreducible p-p interaction $\gam$  
by performing the tranformation of the matrix elements of the interaction
from the p-h to the p-p channel. 
However, this is not enough, since in the p-p channel
a set of additional diagrams must be added arising from p-h 
diagrams which are reducible in that representation.
After including these terms, the interaction contains both direct and exchange
terms, which guarantees antisymmetry and Landau sum rules and, in addition,
it should simultaneously make the nuclear matter Landau parameter ${F}_0$ less 
negative so that the stability condition is satisfied. 
The first contributions to the p-p interaction are depicted in 
line (b) of Fig.~\ref{f:pol}. 
We stress once more that they describe the 
influence of the medium polarization on the nucleon-nucleon interaction within 
the induced interaction model of Babu-Brown.

We discuss now the effects of medium polarization on the superfluidity of 
neutron stars in the channel $\ss$. 
First of all, the irreducible p-p interaction to be used in the 
pairing problem must not include any ladder sum already included in the gap 
equation, 
and therefore the first term in line (b) of Fig.~\ref{f:pol}
is the bare neutron-neutron interaction $V$. 
The next terms include the irreducible p-h interaction in the vertices 
of the p-h bubble. 
If their momentum dependence is neglected, these 
vertices can be identified with the Landau parameters \cite{BACK73}.  

In neutron matter the polarization of the medium is due to density fluctuations
and spin-density fluctuations given by
\bsa
 \delta\rho_{\kv} &=& 
 \delta\rho_{{\kv}\uparrow} + \delta\rho_{{\kv}\downarrow} \:, 
\\
 \delta\rho_{\kv} &=& 
 \delta\rho_{{\kv}\uparrow} - \delta\rho_{{\kv}\downarrow} \:,
\esa
respectively.
Solving the Babu-Brown self-consistent equation, Eq.~(\ref{e:wind}), with 
the $G$-matrix as direct interaction and the renormalized RPA series as 
induced interaction, one determines the p-h interaction and 
eventually, after the Landau angle expansion, the lowest-order Landau 
parameters $F_0$, related to the nuclear compression modulus, and $G_0$,
related to the spin waves.
(In the following the interaction will be expressed
in terms of these two Landau parameters for simplicity). 
Then the effective interaction is calculated including the 
diagrams of Fig.~\ref{f:pol}(b).
The pairing interaction in the $\ss$ channel is then given by    
\be
 \Gamma_{\ss} = V_{\ss} + {\frac{1}{2k_F^2}} \int_0^{2k_F} \!\!\D q \,q 
 \left[ {F_0^2 \uPi(q)\over{1-F_0 \uPi(q)}} -
 {3 G_0^2 \uPi(q)\over{1-G_0 \uPi(q)}} \right]  \:.    
\label{e:pollan}
\ee
This equation shows that the medium screening effect is determined by the 
competition between the attractive term induced by density fluctuations
and the repulsive term induced by spin-density fluctuations (first and 
second term in the bracket, respectively).

\begin{figure}[t] %------------------------------------------------------------
\includegraphics[height=.32\textheight,angle=90,bb=70 350 500 700]
{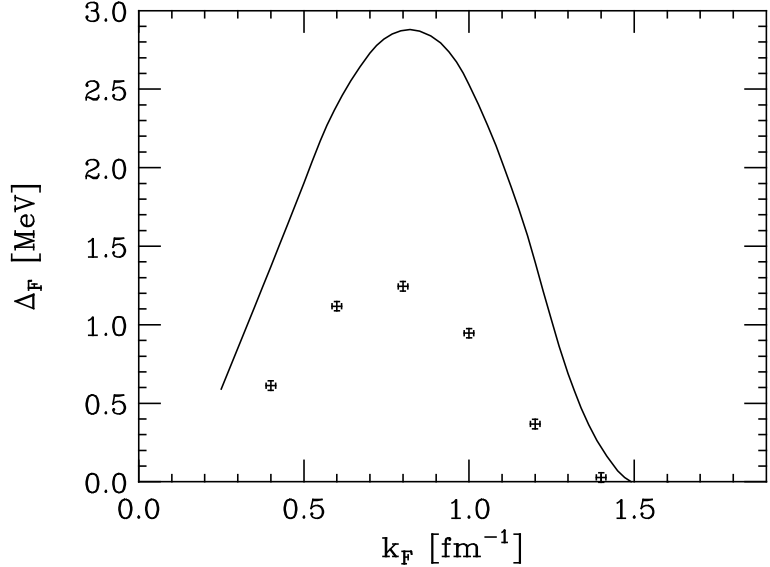}
\caption[]{
$\ss$ pairing gap in neutron matter as a function of the Fermi momentum $k_F$.
The full curve corresponds to using the bare $V_{14}$ potential as direct 
interaction.
The symbols show the effect of the medium polarization described in terms of 
the Landau parameters, according to Eq.~(\ref{e:pollan})}
\label{f:landgap}  
\end{figure} %-----------------------------------------------------------------
 
As depicted in Fig.~\ref{f:landgap}, the effect of the medium polarization 
is an overall suppression of the gap due to the prevalence of the spin-density 
fluctuations over the density fluctuations. 
A similar effect is found in a less crude calculation \cite{SCHU}, 
as shown in Fig.~\ref{f:polgap},
where the peak value is shifted to higher density. 
Such a suppression is common to all calculations existing in the literature 
even if, as to its magnitude, the different predictions do not agree with 
each other, as shown in Fig.~\ref{f:polgap}. 

%==============================================================================
\subsection{Self-energy Effects}
\label{s:self}

Dynamical effects of the interaction on the gap function have been completely 
left aside from the discussion on the medium polarization. 
So far no solution of the gap equation has been attempted considering the 
irreducible interaction block as an energy-dependent quantity. 
On the other hand the energy dependence
in the self-energy can affect deeply the magnitude of the energy gap in a 
strongly correlated Fermi system such as nucleon matter \cite{BOZ,BAL}.  

To discuss self-energy effects we come back to the generalized gap equation 
presented in section~\ref{s:green}. Let us rewrite Eq.~(\ref{e:sgb}) in the 
following form (for the $\ss$ channel)
\be
 \Delta_{k}(\om) = 
 -\int\!\frac{\D^3 k'}{(2\pi)^3} \int\!\frac{\D\om'}{2\pi \I} 
 \,\Gamma_{k,k'}(\om,\om')  { \de_{k'}(\om') \over D_{k'}(\om') }  \:,
\label{e:gge}
\ee
with [cf.~Eqs.~(\ref{e:gg}) and (\ref{e:dd})]
\be
 -D_{k}(\om) = \left[ G_{k}(-\om)G_{k}^{s}(\om) \right]^{-1} =
 G_{k}^{-1}(\om) G_{k}^{-1}(-\om) + \de_k^2(\om)  \:.
\label{e:GAM} 
\ee
The functions $G_{k}(\om)$ and $G_{k}^{s}(\om)$ are the nucleon 
propagators of neutron matter in the normal state and in the superfluid state, 
respectively. 
The $\omega$-symmetry in the two propagators is
to be traced to the time-reversal invariance of the Cooper pairs.
The effective interaction $\Gamma$ is the block of all irreducible diagrams 
of the interaction.
Since we want to focus only on the self-energy effects, we assume the    
interaction to be the bare interaction $V_{k,k'}$, as in BCS.
Then the pairing gap does not depend on the energy (static limit), i.e., 
$\de_k(\om)\equiv \de_k$, and the $\om$-integration can be performed 
in Eq.~(\ref{e:gge}), once the self-energy has been determined. 
A general discussion of the analytic $\omega$-integration of the gap equation 
has been is given in Ref.~\cite{BAL}. 
Here we follow a simplified treatment 
based on the fact that, at each momentum $k$, the main contribution to
the $\omega$-integration comes from the pole of $G_{k}(\om)$. 
This latter is the solution of the implicit equation
\be
 \om_k = k^2\!/2m + \Sigma_k(\om_k) - \mu  \:.
\label{e:pole}
\ee
Expanding the self-energy around the pole $\omega_k$ amounts to expanding
$G^{-1}$ itself, which yields  
\be
 G^{-1}_k(\om) \approx 
 \left( 1 - \frac{\partial\sig}{\partial\om} \Big|_{\om=\om_k} 
 \right)(\om - \om_k)  \:,
\ee
where the prefactor on the r.h.s.~is the inverse  
of the quasiparticle strength $Z_k$ that will be discussed later.
Then the energy dependence of the kernel $D^{-1}_k(\om)$ 
%in Eq.~(\ref{e:gge}) 
takes the simple form
\be
 %D_k(\om) = Z_k^{-2} (\om^2 - \om_k^2 - \Delta_k^2)
 D^{-1}_k(\om) = { Z_k^2 \over \om^2 - \om_k^2 - (Z_k\Delta_k)^2 }  \:,
\ee 
and the $\om$-integration can be performed in the usual way,
leading to 
%After integration one ends up with the equation
\be
 \de_{k} = - \int\! \frac{\D^3 k'}{(2\pi)^3} 
 \left( Z_k V_{k,k'} Z_{k'} \right) 
 \frac{ %{Z}_{k'}^2
 \de_{k'}}{2\sqrt{\om_{k'}^2 + \de_{k'}^2}}  \:.
\label{e:GE}
\ee
Comparing with the BCS result, Eq.~(\ref{e:gap1}), the new gap equation 
contains the quasiparticle strength $Z$ to the second power. 
Since $Z$  
is significantly less than one in a strongly correlated Fermi system, 
a substantial suppression of the energy gap is to be expected.  

Moreover $Z$ deviates from unity only in a narrow region around the Fermi
surface and hence it is not a severe approximation  
to restrict the $\om$-integration to only the pole part at the Fermi energy.  
Expanding then $\Sigma_k(\om)$ around the Fermi surface 
($\om=0$ and $k=k_F$),
Eq.~(\ref{e:pole}) is easily solved and we get 
$\om_k \approx (k^2 - k_F^2)/2m^*$,
where $m^*$ is the effective mass at $k_F$ [see Eq.~(\ref{e:qpe})]. 
The gap equation becomes
\be
 \de_k = - Z_F^2 \int\!{\frac{\D^3 k'}{(2\pi)^3}} \,\frac{V_{k,k'}\de_{k'}}
 {2\sqrt{[(k'^2-k_F^2)/2m^*]^2 + \de_{k'}^2}}  \:,
\label{e:ZZ1}
\ee
where $Z_F^2$ is the quasiparticle strength at the Fermi surface.

As is well known the pairing modifies the chemical potential which is 
calculated self-consistently with the gap equation from the closure equation
for the average density of neutrons
\bea 
 \rho &=& 2\int\!\frac{\D^3 k}{(2\pi)^3} \int\!\frac{\D\om}{2\pi \I} 
 \, G_k^{s}(\om^+)
\\
 &\approx& Z_F \int\!\frac{\D^3 k}{(2\pi)^3} 
 \left[ 1 - \frac{(k^2-k_F^2)/2m^*}{\sqrt{[(k^2-k_F^2)/2m^*]^2 
 + Z_F^2 \de_k^2}} \right]  \:.
\label{e:ZZ2}
\eea
The latter approximation is sufficient to investigate the self-energy effects. 

Before we present the predictions based on Eq.~(\ref{e:ZZ1}),
we discuss some properties of the self-energy $\sig_k(\om)$ of neutron matter.
In the Brueckner approach \cite{MAH} the perturbative expansion 
of $\sig$ can be recast according to the number of hole lines as follows  
\be
 \sig_k(\om) = \sig_k^{(1)}(\om) + \sig_k^{(2)}(\om) + \dots  \:. 
\label{e:SIG}
\ee
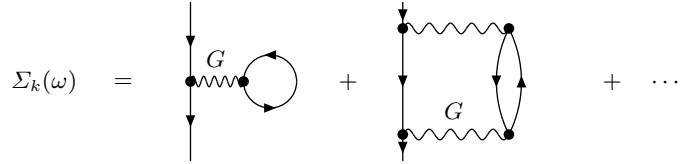
\begin{figure}[h] %------------------------------------------------------------
\begin{picture}(200,50)(0,30)
\Vertex(110,70){2} \Vertex(130,70){2} \Photon(110,70)(130,70){2}{5}
\ArrowLine(110,70)(110,40) \ArrowLine(110,100)(110,70)
\ArrowArc(140,70)(10,0,180) \ArrowArc(140,70)(10,180,360)
\Vertex(190,50){2} \Vertex(190,90){2}
\Vertex(230,50){2} \Vertex(230,90){2}
\ArrowLine(190,100)(190,90) \ArrowLine(190,90)(190,50)
\ArrowLine(190,50)(190,40)
\Photon(190,50)(230,50){2}{5} \Photon(190,90)(230,90){2}{5}
\ArrowArc(270,70)(44.72,153.43,206.56) \ArrowArc(190,70)(44.72,333.43,386.56)
\Text(55,70)[]{$\Sigma_k(\omega)$}
\Text(85,70)[]{$=$} \Text(170,70)[]{$+$} \Text(270,70)[]{$+$}
\Text(290,70)[]{$\cdots$}
\Text(120,79)[]{$G$} \Text(210,59)[]{$G$}
\end{picture}
\caption[]{Hole-line expansion of the self-energy}
\label{f:HL}
\end{figure} %-----------------------------------------------------------------

\begin{figure}[b] %------------------------------------------------------------
\includegraphics[height=.32\textheight,angle=90,bb=130 390 440 750]
{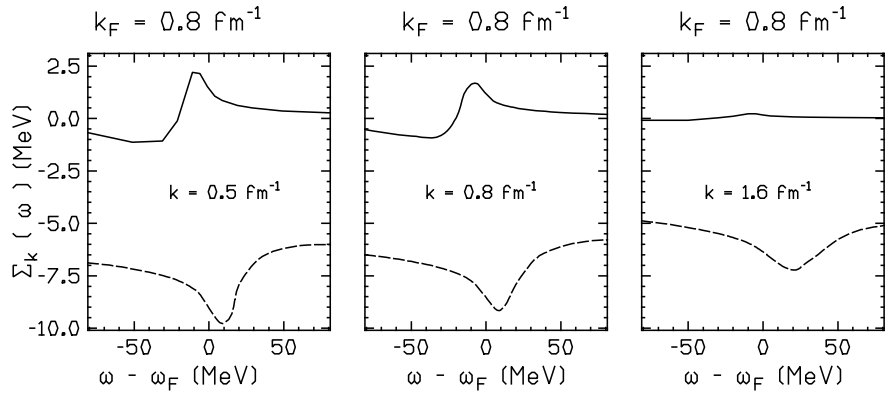}
\caption[]{
Off-shell self-energy in neutron matter at $k_F=0.8\,\rm fm^{-1}$. 
{\em Solid curves}: BHF approximation $\Sigma^{(1)}$. 
{\em Dashed curves}: rearrangement contribution $\Sigma^{(2)}$. 
The Argonne $V_{14}$ potential \cite{V14} has been used 
in the Brueckner calculations}
\label{f:offs}
\end{figure}

\vskip-5mm\noindent
The on-shell values of $\Sigma^{(1)}$ represent the Brueckner-Hartree-Fock 
(BHF) mean field (first diagram in Fig.~\ref{f:HL}); 
the ones of $\Sigma^{(2)}$ represent the so-called rearrangement term 
(second diagram in Fig.~\ref{f:HL}),
which is the largest contribution due to ground-state correlations.
The off-shell values of the self-energy are required to solve the generalized 
gap equation. 
Fig.~\ref{f:offs} displays a typical result for the off-shell neutron 
self-energy $\sig_k(\om)$ calculated up to the second order of the 
hole-line expansion. 
 %-----------------------------------------------------------------
The calculations are based on Brueckner theory adopting 
the continuous choice as auxiliary potential \cite{ZUO}.

Since we are interested in the behaviour of the self-energy 
around the Fermi surface ($k=k_F$ and $\omega = 0$), we may use the expansion
\be
 \sig_k(\om) \approx \sig_{k_F}(0) + 
 \frac{\partial\sig}{\partial\om}\bigg|_F \om + 
 \frac{\partial\sig}{\partial k}\bigg|_F (k - k_F)  \:.
\label{e:dev}    
\ee 
In this case the quasiparticle energy takes the simple form
\be
 \om_k \approx \frac{k^2 - k^2_F}{2m}  
 %\frac{1 + \frac{m}{k_F}\!\left.\frac{\partial\sig}{\partial k}\right|_F}
 %{1 - \left.\frac{\partial\sig}{\partial\om}\right|_F}
 \frac{1 + (m/k_F)(\partial\sig/\partial k)|_F}
      {1 - (\partial\sig/\partial\om)|_F}
 = \frac{k^2 - k^2_F}{2 m^*}  \:.
\label{e:qpe}
\ee
In the previous equation we have introduced the effective mass $m^*/m$ 
as the product of the $e$-mass $m_e$ and the $k$-mass $m_k$, 
which are defined respectively as follows \cite{MAH}
\bsa
 {m_e\over m} &=&  1 - \frac{\partial\sig}{\partial\om}\Big|_F  
 = {1\over Z_F}  \:,
\label{e:ME}
\\
 {m_k\over m } &=& 
 \left[ 1 + \frac{m}{k_F} \frac{\partial\sig}{\partial k}\Big|_F \right]^{-1} 
 \:.
\label{e:MP}
\esa 
The partial derivatives are evaluated at the Fermi surface.
The $k$-mass is related to the non-locality of the mean field and, 
in the static limit ($\om=\om_F$), coincides with the effective mass. 
This quantity is of great interest in heavy-ion collision physics,  
since the transverse flows are very sensitive to the momentum dependence  
of the mean field. 
The $e$-mass is related to the quasi-particle strength. 
This latter gives the discontinuity of the neutron momentum distribution 
at the Fermi surface, and measures the amount of correlations included in 
the adopted approximation.

\begin{figure}[b] %------------------------------------------------------------
\includegraphics[height=.32\textheight,angle=90,bb=30 390 490 750]
{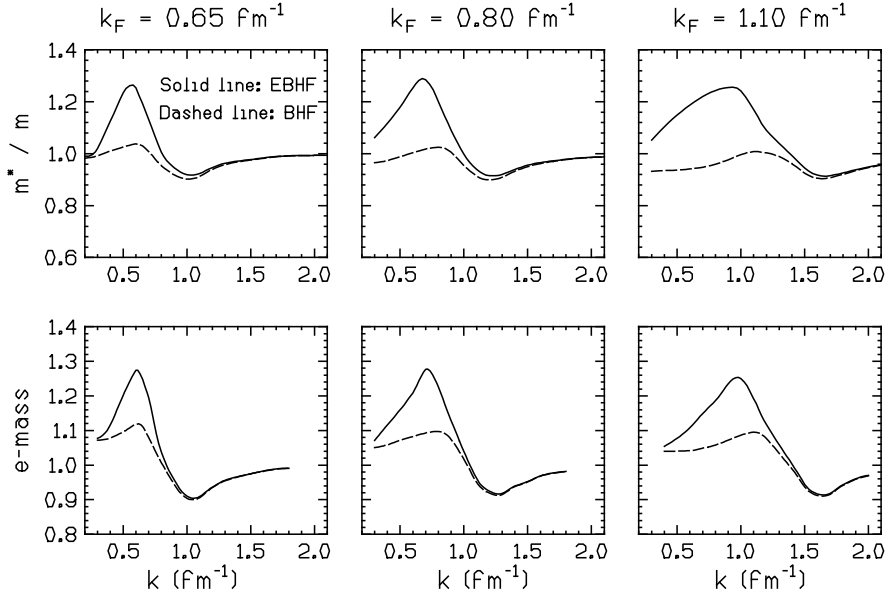}
\caption[]
{Effective masses in neutron matter for three densities: 
 effective mass $m^*$ (upper panels) and $e$-mass (lower panels). 
 Dashed lines correspond to $\sig^{(1)}$ and solid lines to 
 $\sig^{(1)}+\sig^{(2)}$}
\label{f:OFFMS}
\end{figure} %-----------------------------------------------------------------

From $\sig_k(\omega)$ the effective masses  
are extracted according to Eqs.~(\ref{e:ME}) and (\ref{e:MP}). 
They are depicted in Fig.~\ref{f:OFFMS}, where the full calculation is 
compared to that including only the BHF self-energy.
We may distinguish two momentum intervals: 
at $k\approx k_F$ the momentum dependence of the effective mass $m^*$ is 
characterized by a bump, whose peak value exceeds the value of the bare mass; 
far above $k_F$ the bare mass limit is approached. 
The contribution from the rearrangement term exhibits a
pronounced enhancement in the vicinity of the Fermi energy, which is to be
traced back to the high probability amplitude for p-h excitations 
near $\epsilon_F$ \cite{ZUO}. 
At high momenta this contribution vanishes. 
One should take into account that in this range of $k_F$ the neutron
density is quite small 
($\rho=0.074\,\rm fm^{-3}$ at the maximum $k_F=1.3\,\rm fm^{-1}$). 
This behaviour of the effective mass $m^*$ is mostly due to the $e$-mass, 
as shown in the lower panels of Fig.~\ref{f:OFFMS}. 
In all panels of Fig.~\ref{f:OFFMS} is also reported 
for comparison the effective mass in the BHF limit 
(only $\Sigma^{(1)}$ included), 
which exibits a much less pronounced bump at the Fermi energy.

\begin{figure}[b] %------------------------------------------------------------
\includegraphics[height=.32\textheight,angle=90,bb=70 350 500 700]
{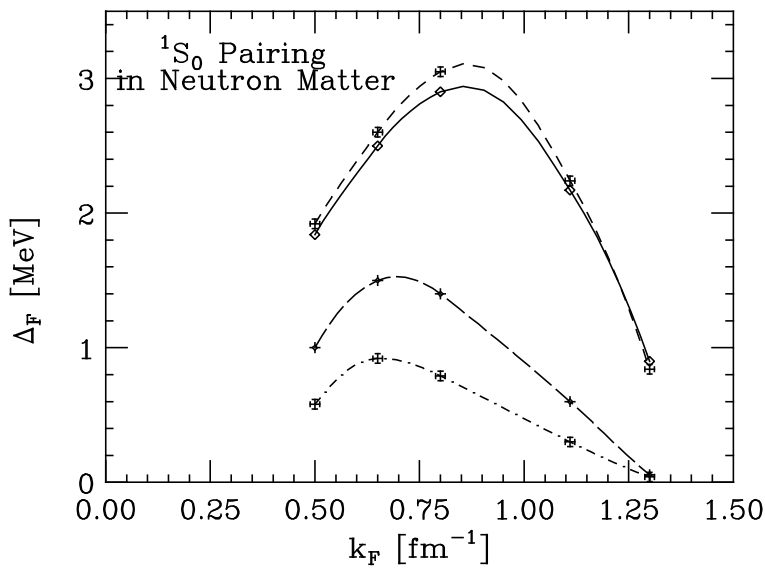}
\caption[]
{Energy gap in different approximations for the self-energy: 
free s.p.~spectrum ({\it solid line}); 
effective mass with $Z_F=1$ ({\it upper dashed line});
$Z$ from $\Sigma^{(1)}$ ({\it long dashed line}); 
$Z$ from $\Sigma^{(1)}+\Sigma^{(2)}$ ({\it lower dashed line})}
\label{f:selfen}
\end{figure}

With the off-shell values of the self-energy discussed above as input,
the gap equation has been solved in the form of Eq.~(\ref{e:ZZ1}), 
coupled with Eq.~(\ref{e:ZZ2}) \cite{SCK}.
This is a quite satisfactory approximation, especially in
view of studying the self-energy effects on pairing. 
The Argonne $V_{14}$ potential has been adopted as pairing interaction, 
which is consistent
with the self-energy data for which the same force has been used.
The results are reported in Fig.~\ref{f:selfen} for a set of different 
$k_F$-values. 
 %-----------------------------------------------------------------
The diamonds connected by a solid line represent the solution
of Eq.~(\ref{e:ZZ1}) replacing the effective mass by the free one, 
$m^*=m$, together with $Z=1$. 
This is very close to the prediction obtained from the BHF
approximation for the effective mass $m^*$, but still keeping $Z=1$
(fancy crosses connected by the dashed line).
This similarity stems from the fact that at the Fermi surface
$m^*/m$ from BHF is close to one, as shown in Fig.~\ref{f:OFFMS}.
The self-energy effects are estimated in two approximations. 
In the first one $m^*$ and the $Z$-factor are calculated from a BHF code. 
In the considered density domain the $Z$-factor ranges around $0.9$.
Despite its moderate reduction a dramatic suppression of the gap is
obtained, as shown by the long dashed line in Fig.~\ref{f:selfen}.
It is due to the exponential dependence of the gap on all quantities.
Still a further but more moderate reduction is
obtained when the rearrangement term is included in the second approximation. 
The smaller $Z$-factor ($Z\approx 0.83$ at $k_F=0.8\,\rm fm^{-1}$)
is to a certain extent counterbalanced by an increase of the effective mass 
($m^*/m \approx 1.2$ at the same $k_F$).

From the previous discussion we may conclude that including self-energy 
effects is an important step forward in understanding the pairing in nucleon 
matter, but one should also include, on an equal footing, vertex corrections.

%==============================================================================
\section{Conclusions}

This chapter addressed the present status of the theoretical progress on
the superfluidity of neutron matter and the microscopic calculations
of nucleonic pairing gaps relevant for the conditions that are encountered
in the interior of a neutron star. 
We have seen that most of the results that can be found in the
literature, are obtained within the BCS approximation.

Unfortunately, as has also been pointed out, this approximation cannot
be considered reliable in any region of density,
since neutron matter is a strongly correlated Fermi system:
The same nucleons that participate to the pairing coupling also participate
to screen the pairs. 
This requires necessarily to improve the interaction kernel by in-medium 
polarization corrections that, at the present time, 
can be treated with confidence only at very low density,   
where the ring series can be truncated at the order of the one-bubble term.

The induced interaction approach is a promising candidate to describe these 
effects, but its predictions are still affected by too severe approximations. 
The existing calculations addressing this problem point to a substantial
reduction of pairing in the ($T=1$) $\ss$ channel with respect to the BCS
results.
In the $\pf$ channel, no estimate of polarization effects has been made so
far, at all.

Also self-energy effects, when properly included in the gap equation,  
strongly influence the pairing mechanism. 
We saw in fact that the gap is reduced with the square of the 
quasi-particle strength $Z$.    

In any case it remains the problem of a complete solution of the generalized
gap equation, taking into account simultaneously screening and self-energy
corrections, which moreover have to be treated on the same footing.
This latter goal remains a considerable theoretical challenge for the future.

Due to lack of space and time we have not discussed more speculative 
subjects like 
pairing in isospin asymmetric matter,
relativistic effects on pairing, 
hyperon pairing, etc., 
which might also have relevance for neutron star physics.

%==============================================================================
%\section*{Acknowledgements}

%%%%%%%%%%%%%%%%%%%%%%%%%%%%%%%%%%%%%%%%%%%%%%%%%%%%%%%%%%%%%%%%%%%%%%%%%%%%%%%

%INDEX%%%%%%%%%%%%%%%%%%%%%%%%%%%%%%%%%%%%%%%%%%%%%%%%%%%%%%%%%%%%%%%%%%%%%%%%%
% Please check with the editor of your book whether he plans to
% include a "mutual" subject index - if so, please code your entries
% in the standard syntax. For your own purposes you may print your
% "personal" index by using the following commands:
%
%\clearpage
%\addcontentsline{toc}{section}{Index}
%\flushbottom
%\printindex
%%%%%%%%%%%%%%%%%%%%%%%%%%%%%%%%%%%%%%%%%%%%%%%%%%%%%%%%%%%%%%%%%%%%%%%%%%%%%%%

\end{document}